# Best Practices for Data Publication in the Astronomical Literature

Tracy X. Chen ,[1] Marion Schmitz ,[1] Joseph M. Mazzarella ,[1] Xiuqin Wu,[1] Julian C. van Eyken ,[2]
Alberto Accomazzi ,[3] Rachel L. Akeson ,[2] Mark Allen ,[4] Rachael Beaton ,[5] G. Bruce Berriman ,[2]
Andrew W. Boyle ,[2] Marianne Brouty,[4] Ben H. P. Chan,[1] Jessie L. Christiansen ,[2] David R. Ciardi ,[2]
David Cook ,[1] Raffaele D'Abrusco ,[3] Rick Ebert ,[1] Cren Frayer,[1] Benjamin J. Fulton ,[2]
Christopher Gelino,[2] George Helou ,[1] Calen B. Henderson ,[2] Justin Howell ,[6] Joyce Kim,[1]
Gilles Landais ,[4] Tak Lo,[1] Cecile Loup,[4] Barry Madore ,[7,8] Giacomo Monari ,[4] August Muench ,[9]
Anais Oberto ,[4] Pierre Ocvirk ,[4] Joshua E. G. Peek ,[10,11] Emmanuelle Perret,[4] Olga Pevunova,[1]
Solange V. Ramirez,[7] Luisa Rebull ,[6] Ohad Shemmer ,[12] Alan Smale ,[13] Raymond Tam,[2] Scott Terek,[1]
Doug Van Orsow ,[13,14] Patricia Vannier,[4] and Shin-Ywan Wang[1]

[1] Caltech/IPAC-NED, Mail Code 100-22, Caltech, 1200 E. California Blvd., Pasadena, CA 91125, USA
[2] Caltech/IPAC-NExScI, Mail Code 100-22, Caltech, 1200 E. California Blvd., Pasadena, CA 91125, USA
[3] Center for Astrophysics | Harvard & Smithsonian, 60 Garden Street, Cambridge, MA 02138, USA
[4] Centre de Données astronomiques de Strasbourg, Observatoire de Strasbourg, 11, rue de l'Université, 67000 STRASBOURG, France
[5] Department of Astrophysical Sciences, Princeton University, 4 Ivy Lane, Princeton, NJ 08544, USA
[6] Caltech/IPAC-IRSA, Mail Code 100-22, Caltech, 1200 E. California Blvd., Pasadena, CA 91125, USA
[7] The Observatories, Carnegie Institution for Science, 813 Santa Barbara Street, Pasadena, CA 91101, USA
[8] Department of Astronomy & Astrophysics, University of Chicago, 5640 S. Ellis Avenue, Chicago, IL 60637, USA
[9] American Astronomical Society, 1667 K Street NW, Suite 800, Washington, DC 20006, USA
[10] Space Telescope Science Institute, 3700 San Martin Drive, Baltimore, MD 21218, USA
[11] Department of Physics & Astronomy, Johns Hopkins University, 3400 N. Charles Street, Baltimore, MD 21218, USA
[12] Department of Physics, University of North Texas, Denton, TX 76203, USA
[13] NASA Goddard Space Flight Center, 8800 Greenbelt Road, Greenbelt, MD 20771, USA
[14] Business Integra, Inc., 6550 Rock Spring Dr #600, Bethesda, MD 20817, USA

## ABSTRACT

We present an overview of best practices for publishing data in astronomy and astrophysics journals. These recommendations are intended as a reference for authors to help prepare and publish data in a way that will better represent and support science results, enable better data sharing, improve reproducibility, and enhance the reusability of data. Observance of these guidelines will also help to streamline the extraction, preservation, integration and cross-linking of valuable data from astrophysics literature into major astronomical databases, and consequently facilitate new modes of science discovery that will better exploit the vast quantities of panchromatic and multi-dimensional data associated with the literature. We encourage authors, journal editors, referees, and publishers to implement the best practices reviewed here, as well as related recommendations from international astronomical organizations such as the International Astronomical Union (IAU) for publication of nomenclature, data, and metadata. A convenient *Checklist of Recommendations for Publishing Data in the Literature* (Appendix A) is included for authors to consult before the submission of the final version of their journal articles and associated data files. We recommend that publishers of journals in astronomy and astrophysics incorporate a link to this document in their Instructions to Authors.

*Keywords:* Astronomical Reference materials (90), Astronomy databases (83), Astronomical object identification (87), Astrometry (80), Astronomical coordinate systems (82), Photometry (1234), Observatories (1147), Telescopes (1689)

## 1. INTRODUCTION

Modern telescopes and instruments are being used to observe larger areas of the sky over wider ranges of the spectrum, and with greater frequency than ever before. The volume and complexity of resulting data are growing at an exponential rate, not only within the astronomical archives that capture the original data products, but also in the scientific literature where more highly-processed data are published alongside scientific results. It is therefore extremely important that data appearing in journal articles are prepared and published with great diligence in order to: accurately present new data, avoid any loss of information, preserve and support the integrity of scientific results, and enhance the reusability of data to facilitate further analysis and followup studies. The impacts of publishing complete and accurate data are far-reaching across all astrophysical disciplines and encompass all types of objects, and are important to the scientific process for providing transparency and reproducibility in the primary record of scientific exploration and discovery.

These data also serve another vital function in modern science as they are continuously being integrated into astronomical archives to provide comprehensive information for astronomical objects, and to support planning new observations, performing data analysis, making new discoveries, and preparing publications of new results. Uniformity and adherence to established norms in published data are essential to improving automation, efficiency, and accuracy of procedures required to integrate data from the literature into astronomical archives in a timely fashion. The high degree of connectivity between the digital journal articles and the astronomical archives, with the NASA Astrophysics Data System (ADS)[1] as a nexus through its extensive bibliographic database, provides an ecosystem used 24/7 by thousands of scientists around the world who depend on astronomical data that are as current and accurate as possible.

Much of data appearing in the astronomical literature are presented with great care and serve two scientific needs well: 1) a high quality scientific record with results that can be reproduced and expanded upon in follow-up studies, and 2) clean data that can be straightforwardly integrated into astronomical databases. However, there are a substantial number of journal articles published each year where various issues continue to severely detract from the quality and utility of the data and hinder both scientific goals. The most common and severe example is the publication of ambiguous object names, typically truncated coordinate-based names, that make it nearly impossible to reproduce observations or accurately cross-identify these sources with those in other journal articles or catalogs. Other common examples include publishing data without uncertainties or with an unrealistic number of significant figures, not specifying the reference frame for coordinate or redshift measurements, and placing data critical to (re)producing science results in a personal URL that has no long-term access support. One study has shown that in 2011, 44% of data links published in the astronomical literature a decade earlier (in 2001) were broken (Pepe et al. 2014). Most of these issues, and more, can be avoided or minimized with a small amount of additional effort on the part of authors, combined with more attention to flagging such issues by referees and editors so authors can correct errors and omissions before the article is published. To assist in this process, we provide in this article recommendations on best practices for publishing data and metadata in astronomical journals. The topics discussed here are essential to achieving the goals of data sharing, open access, and reproducibility of science results, many of which also reflect the FAIR (Findable, Accessible, Interoperable, and Reusable) guiding principles for scientific data management and stewardship (Wilkinson et al. 2016).

A number of issues discussed here have also been addressed by Cambrésy et al. (2011) in the context of the SIMBAD (Wenger et al. 2000; SIMBAD Team 2022a)[2] and VizieR services (Ochsenbein 1996; Ochsenbein et al. 2000)[3] of CDS.[4] We encourage all authors, referees, editors and publishers to follow these best practices during the preparation, submission, refereeing and editing stages of the publication process. We expect that these best practices will evolve as new forms of data are published in the journals, and as the journals and archives receive feedback from authors. When enough revisions build up, we plan to publish updates to this original publication on arXiv.[5] The broader topic of standards, formats and best practices for depositing specific types of data products in repositories and archives is beyond the scope of this article.

## 2. ASTROPHYSICAL DATA GUIDELINES

Here we provide guidelines for presenting data in many areas of astrophysical research: stars, galaxies, dust and gas, and planets. Names, astrometry, photometry, redshift/velocity etc. are among the fundamentals of the

---

data published in an astrophysical article. Formatting and referencing these data correctly are essential in understanding the data itself and the derived science results. Before addressing each of these categories individually, here are some general rules:

- **Define symbols, acronyms, and abbreviations.** Symbols, acronyms, and abbreviations must be clearly defined when used anywhere in a publication, generally at first use, even if they are common throughout the discipline. For example, "... used by the Dark Energy Survey (DES) ..." This is especially important in younger fields of investigation where symbols may be used in divergent ways by the community such that their meaning is not as clear is it might seem. If there is any doubt, define everything explicitly, or at the least cite a source which clearly defines the conventions adopted. The definition should be consistent through the entire article to avoid confusion.

- **Provide uncertainty and confidence level when reporting a new measurement.** Avoid using parentheses after the last digits of the measured value to indicate uncertainty, as this is very unclear with a high degree of probable confusion. For example, the period of a periodic phenomenon should be given as "P = 1.23456±0.00012 days" instead of "P = 1.23456(12) days". The uncertainty in the latter representation could be misinterpreted as "±0.0000012 days".

- **Present the appropriate number of significant figures for numerical measurements and uncertainties.** Significant figures in a measurement indicate its precision. When data are generated from floating-point machine computation, there is a tendency to display more significant figures than are justified by the measurement. A decimal degree coordinate of (131.32134587°, 1.01243229°) would imply an accuracy of $10^{-8}$ degrees (or $0.00001''$), which is not obtainable by most current telescopes and thus over-represent the precision of the measurements. Attention is also needed to coordinate the number of decimal places in the measurements and associated uncertainties. For example, present a measurement as 0.123±0.002, not 0.12345±0.002 or 0.123±0.00234.

- **Report the units for measurements if present, and adopt commonly-used ones.** This will ensure the correct representation of a physical quantity, and facilitate the comparison between results from various observations, models, or analyses. For example, present source color "(B-V) = 0.45 mag" instead of "(B-V) = 0.45". Please refer to Rots et al. (2021) for recommendations by the International Astronomical Union (IAU)[6] on units in astronomy and astrophysics.

- **Indicate preferred values if applicable.** When reporting multiple measurements for one parameter or alternative parameter sets from different techniques or algorithms that fit observations, it is valuable to the readers and beneficial to the archives that the authors indicate the preferred value or parameter set. For example, Grieves et al. (2021) provided multiple solutions for NGTS-13 stellar and companion parameters, and the preferred solution was clearly stated in the text and indicated in bold in Table 4 of the article.

## 2.1. Nomenclature

The most common and basic type of data in any observational article is the naming of astronomical sources. Unambiguous names that follow recommended nomenclature standards are essential prerequisites for clear communication of observational results and scientific conclusions, and for ensuring follow-up observations target the proper source.

### 2.1.1. IAU conventions

When assigning object names in an article, we recommend the authors follow the established conventions from the IAU such as "Naming of Astronomical Objects" (International Astronomical Union n.d), "How to refer to a source or designate a new one" (IAU Working Group on Designations 2018a), and "Specifications concerning designations for astronomical radiation sources outside the solar system" (IAU Working Group on Designations 2018b). The IAU Dictionary of Nomenclature of Celestial Objects (Lortet et al. 1994; SIMBAD Team 2022b) provides a list of acronyms that are currently in use, and should be consulted to confirm the correct acronyms and formats for known objects, and to avoid reusing the same ones for newly discovered objects. When publishing simultaneous independent discoveries of the same astrophysical objects, reasonable efforts should be made to coordinate and avoid publishing conflicting designations.

Following the IAU guidelines, we recommend:

- **Provide the complete object name.** A name that has the coordinate part truncated will likely

---

[6] https://www.iau.org/

be ambiguous and can be confused with a nearby object (see, e.g., first example in Table 1).

- **Explicitly include the "J" in names based on J2000 coordinates.** Without the "J", this could be misinterpreted as B1950, resulting in an incorrect object position and telescope pointing. For example, use "BR J0529-3526" instead of "BR 0529-3526".

- **Insert spacers between a catalog name and the identifiers within the catalog.** For example, use B3 2327+391, not B32327+391. This will prevent mixing and misinterpreting the catalog name and the identifiers, and allow both the readers and name resolver services at the archives (e.g., SIMBAD, NED,[7] NASA Exoplanet Archive[8]) to recognize object names more efficiently.

- **Distinguish between part of an object and the object itself.** For example, use "3C 295 cluster" instead of "3C 295" when referring to the cluster. Similarly, the hosts for transients, like supernovae and gamma-ray bursts, should be referred to with the proper names when host properties are discussed, instead of just using the names of the transients.

- **Do not use the same name for different objects.** Once a name has been assigned to an object in a published catalog or journal article, it should not be reused for a different source in the future, even if an object's existence is refuted. For example, the tau Ceti system now has four planets: e, f, g, and h. Since tau Ceti b, c, and d were refuted, the letter designations b, c, and d were not reused for the newer planets to avoid confusion.

Table 1 illustrates some ambiguous/improper astronomical designations that have appeared in the literature, along with the recommended proper usage.

### 2.1.2. *New objects*

For newly discovered objects, please follow the IAU conventions mentioned in §2.1.1 for designations. Additional recommendations include:

- **Confirm the object is new.** Before calling a discovery, please check in established astronomy

databases[9] to see if there is indeed no prior reported detection at the location. For example, one can do a Near Position search[10] in NED for an extragalactic object. We note that this step only applies to studies of a small number of objects. For large survey catalogs, new identifiers are usually assigned without confirming if every detection is new.

- **Assign a name.** Without a proper name, it is difficult for both the readers and the databases to unambiguously reference information in a publication. It is recommended to submit any new acronym to the IAU Working Group on Designations and Nomenclature[11] for review and registration.

- **Verify the name is unique.** The key to object designations is that each one must be unique when compared to objects identified at other observatories and wavelengths, especially within the same article. When creating new acronyms, please consult the IAU Dictionary of Nomenclature of Celestial Objects as noted in §2.1.1 to avoid reusing existing ones.

- **Keep the appropriate number of significant figures in coordinate-based names.** When the name of an object is generated from the coordinates of the object, too many significant figures would imply a much higher accuracy than the measured position of the object. For example, J092712.64+294344.0 indicates a positional accuracy of 0.15 arcsec while J092712.644+294344.02 indicates an accuracy of 0.015 arcsec. Conversely, it is very important to include all significant figures in coordinate-based source names, at least the first time mentioned in the article, because truncating coordinates often leads to ambiguity and difficulty cross-matching the source with prior data due to confusion with nearby objects.

### 2.1.3. *Known objects*

For objects that are already published and known, we recommend:

- **Use established names.** It is unnecessary and often adds confusion to give a new name to an ob-

---

**Table 1**: Examples of improper astronomical designations in literature

| As published | Why it is improper | Recommended usage (notes if available) |
|---|---|---|
| SDSS J1441+0948 | Insufficient precision in RA and DEC can cause confusion with nearby sources. | SDSS J144157.24+094859.1, or SDSS J144156.97+094856.5, or SDSS J144157.26+094853.7 |
| SN 05J | Incomplete name can be interpreted into different objects. | SN 1905J, or SN 2005J |
| HESS J232+202 | Leading zero in RA is missing and can cause misinterpretation of the RA at 23 hours instead of 02 hour. | HESS J0232+202 |
| BR 0529-3526 | Missing letter J to specify J2000 equatorial coordinates. | BR J0529-3526 |
| 0008+006 | Name prefix is needed to distinguish between different objects. | ZC 0008+006 (Redshift z = 2.3), or IVS B0008+006 (Redshift z = 1.5) |
| DEM45 | H II regions in LMC or SMC should be indicated with "L" or "S" to avoid ambiguity. | DEM L 045, or DEM S 045 |
| SDSS 587729386611212320 | Database objectID numbers are used without specifying release number. The same running number may refer to a different source in a different release. | SDSS DR6 587729386611212320 |
| Gaia DR 2 2.7904e18 | ID is written in scientific notation, making it impossible to retrieve the actual object. | Gaia DR2 2790494815860044544 |
| mu cep | Ambiguous name can be interpreted into different objects. | $\mu$ Cep (21h43m30.46s, +58d46m48.2s, ICRS J2000), or MU Cep (22h23m38.63s, +57d40m50.8s, ICRS J2000) |

ject that already has a name. Creating a fanciful name for an object with an existing designation is especially discouraged. Using the established names also gives proper credit to the original authors or survey team that first discovered and cataloged the source.

- **Check for the correct formatting.** For astronomical objects outside the Solar System, authors are encouraged to validate all the identifiers for known objects in their publications through Sesame,[12] a service hosted by CDS that queries NED, SIMBAD, and VizieR to help resolve object names. If some valid names are not indexed by these services, authors should verify the names with the discovery papers of the objects.

#### 2.1.4. *Cross-identifications*

When multiple designations exist for the same object, the different names are cross-identified by authors or databases. The best practices with these cross-identifications are:

- **Confirm the names and positions.** Many examples exist in the literature where multiple columns of cross-identifications disagree with each other or with the listed coordinates. We suggest to always verify with established databases that all of the names given to an object are valid cross-identifications for the object and that the listed positions are for the same object. If differences are found among cross-identifications, we suggest that authors contact the relevant databases.

- **Cross-match the same objects within the same article.** If the same object appears in mul-

---

[12] SIMBAD Team (n.d); https://cds.unistra.fr/cgi-bin/Sesame

tiple tables of the same article, but with different designations, a cross-matching of the tables should be done by the authors to enable the readers to quickly access related data about the objects. For example, Table 4 of Kundu et al. (2007) provided for the same objects both their X-ray identification number as given in Table 2 and optical identification number as in Table 3 of the article, and therefore linked the position and photometry data for the objects discussed in all three tables.

## 2.2. *Astrometry*

One of the primary attributes for an astronomical source is its location in the sky. Accurate celestial coordinates for the objects are especially important for followup observations and study. When presenting coordinates in a publication, we recommend:

- **Provide the best available coordinates.** Precise positions of the sources are indispensable for the usability of observational data and for planning followup observations. All objects studied in an article, especially those from private catalogs, need to be presented with coordinates. Complete celestial coordinates are preferred, e.g., 12h34m56.78s, +12d34m56.7s (Equatorial J2000). When positional offsets are published instead, it is essential that authors include the coordinates of the reference position, as well as the angle of rotation (if North is not up or East is not to the left), and the sense of the offset (i.e., reference point minus source, or source minus reference point).

- **Specify the celestial reference system and/or frame.** Indicate the reference system and/or frame for the coordinates (Urban & Seidelmann 2013, chap. 4 and 7). The current IAU celestial reference system is the International Celestial Reference System (ICRS; Arias et al. 1995; IERS 2013a), and it is realized through the International Celestial Reference Frame (ICRF; Ma et al. 1998; IERS 2013b).

- **Indicate the equinox and epoch of observations when necessary.** This is particularly important for the positions of stars and exoplanet systems. Nearby stars can have a quite significant proper motion (>1 arcsec/year), which means that the position changes with time and therefore requires the equinox and epoch of observation in order to compute its position at another epoch. The standard equinox and epoch currently in use are J2000.0 (Aoki et al. 1983).

- **State the wavelength range from which astrometry is obtained, where appropriate.** Astronomical sources can be detected at slightly different positions at various wavelengths due to different emission mechanisms of the components. Therefore, providing the wavelength range of the detection is important for understanding the various components of the object and for cross-identifications between sources detected at various wavelengths.

We caution that although the standard equinox and epoch currently in use are J2000.0, when citing coordinates from a catalog, a web page or other sources, one cannot assume the equinox/epoch is always J2000.0. For example, the reference epoch for the Gaia Early Data Release 3 is J2016.0, while it is J2015.5 for Gaia Data Release 2 and J2015.0 for Gaia Data Release 1.[13]

## 2.3. *Photometry*

The flux or intensity of light radiated by astronomical objects is another key observable in astronomy. To properly represent the photometry data and to enable easy comparisons of the results, we strongly recommend:

- **State the facility, telescope and instrument used.** Specify whether the facility is ground-based or space-based. Authors also need to provide any other relevant instrument configuration information, the specific camera on the instrument, and/or the specific CCD chips of the camera that were used at the time of the observation. This information is crucial to the proper interpretation of the data, e.g., hardware changes such as the replacement of a filter at a particular time will lead to slightly different instrument responses. Good documentation of the metadata will also facilitate the correct and fast ingestion of photometry data in services such as the NED spectral energy distribution (SED) plots and the VizieR Photometry viewer,[14] which in return will better serve the community.

- **Describe the method used to estimate photometry.** Indicate if an estimate is from point spread function fitting, aperture photometry, isophotal measurements, etc. If it is aperture photometry, report the size of the aperture and background annulus, and any corrections made in the calculation.

---

[13] https://www.cosmos.esa.int/web/gaia/earlydr3
[14] https://vizier.unistra.fr/vizier/sed/

- **Use standard passband/filter identifiers.** Do not modify or abbreviate the identifier as it may conflict with a different standard identifier. For example, indicate "Johnson B" or "Cousins B" instead of just "B"; use "2MASS $K_s$" instead of just "K". A listing of commonly-used identifiers is available at the Spanish Virtual Observatory Filter Profile Service (Rodrigo et al. 2012; Rodrigo & Solano 2020; Spanish Virtual Observatory 2022).[15] When possible, authors should provide a link to the instrument documentation with the actual response curve for the filter in question.

- **Clarify the magnitude system.** Explicitly indicate whether a magnitude is on the AB, Vega, ST, or some other magnitude system. For example, the absolute magnitudes of the Sun in the 2MASS J band, for the AB and ST systems, are 4.54 mag and 6.31 mag, respectively (Willmer 2018).

- **Specify spectral transitions completely.** Describe the molecular species, transitions, and frequencies/wavelengths. For example, carbon monoxide (CO) has several detectable transitions as do $^{13}CO$ and $C^{17}O$. The most commonly observed transition is ($J$=1-0) and each is between 110 and 115 GHz. To clearly define a spectral transition, one should use, e.g., "CO ($J$=1-0) $\nu$ = 115 GHz".

### 2.4. *Time*

Knowing the time of observation is essential to understand the observed properties of many astronomical objects, and often for calculating positions, especially for transient or variable sources and moving objects. Calibrations sometimes may also vary depending on when the instrument was installed on a telescope. Authors are advised to:

- **Provide the time of observation and exposure time.** Any times should be explicitly described in terms of both the frame of reference (e.g., JD, BJD, HJD), and the time system used (e.g., UTC, TDB, TAI). For example, use "BJD-TDB" to indicate Barycentric Julian Date in the Barycentric Dynamical Time standard (preferred). This is particularly important when precise timing is needed, such as the measurement of exoplanet transit timing variations. See Eastman et al. (2010) and Urban & Seidelmann (2013, chap. 3) for a helpful discussion of precise time standards. It is also important to specify the duration (exposure time) and whether the presented observation time is the beginning, midpoint, or end of the exposure time. When a precise time is not meaningful (such as detections from stacked images), a time range where these observations occurred should be provided.

- **Favor full Julian Dates over abbreviated or offset Julian Dates.** When reporting Julian Dates, the full unmodified date (e.g., 2456789.123) is preferred over any offset variation (e.g., 6789.123), to avoid confusion. This also helps archives avoid having to track down and add often arbitrary offsets to put observations on a uniform time scale, which can add an opportunity for errors to be introduced. Where an offset variation must be used, be sure to clearly indicate the value of the offset (e.g., JD-2454833.0) and to refer to the abbreviated date as Modified Julian Date (MJD) **only** when the offset meets the IAU-defined formal definition of Julian date minus 2400000.5; all other abbreviated Julian Dates should be referred to as Reduced Julian Date. The IAU has recommended, "that where there is any possibility of doubt regarding the usage of Modified Julian Date, care be exercised to state its definition specifically" (The XXIIIrd International Astronomical Union General Assembly 1997).

- **Include phase timing measures along with reported periods, where relevant and practical.** This allows for future observations to be phased against your data, and combined. For example, for a transiting exoplanet orbit where the period is known, include a time of transit.

- **State when observations from multiple missions are executed simultaneously.** Coordinated observations in high-energy astronomy are critical for making a simultaneous spectral fit, which can help determine the physical processes responsible for emission in a particular energy band. It is also important to specify which instruments/cameras/chips were taking data and analyzed during these coordinated observations. If possible, include a graphical representation of the times that the missions obtained the data to help visualize where the simultaneity occurs (e.g., Figure 2 of Abbott et al. 2017).

### 2.5. *Redshift/velocity*

---

[15] http://svo2.cab.inta-csic.es/theory/fps/

The redshift or the recessional radial velocity of an object is of major importance to many astrophysical fields. The recommendations on publishing redshift/velocity data include:

- **Describe the method used to obtain redshift.** This includes the particular method (spectroscopic, photometric, Friends-of-Friends, etc.) and base assumptions used in the models (template fitting, machine learning, etc). When available, include a reference to the model/method used to determine the redshift.

- **Specify the reference frame of the redshift measurements.** Be sure to include a clear indication of the reference frame, e.g., heliocentric, barycentric, Galactocentric, or LSR (Local Standard of Rest). Additional definition on the solar velocities adopted and likely also the knowledge of the Sun's distance to the Galactic center is needed for Galactocentric velocities.

- **Provide the frequency/wavelength of the measurement.** This is important due to the fact that the measured redshift (thus the derived radial velocity) measured from different spectral features may represent very different physical phenomena. For example, a redshift measured from H I 21 cm emission line may have a significantly different systematic velocity than a redshift measured from stellar absorption lines in the same galaxy.

- **State the velocity definition (radio or optical) when reporting radial velocity.** Velocity is conventionally defined from the resultant frequency ($\nu$) shifts in radio astronomy, i.e., $v = c(\nu_{rest} - \nu_{obs})/\nu_{rest}$, and defined from the resultant wavelength ($\lambda$) shifts in optical astronomy, i.e., $v = c(\lambda_{obs} - \lambda_{rest})/\lambda_{rest}$ $(= c(\nu_{rest} - \nu_{obs})/\nu_{obs})$. Here the subscripts "obs" and "rest" refer to the observed and the rest frame (i.e., emitted), and c is the speed of light. The radio velocity increment depends upon the rest frequency, whereas the optical velocity increment depends on the observing frequency. Therefore, the velocities defined from these two conventions can differ significantly, especially for large values of v.[16]

- **Indicate the quality of the measurement when possible.** A qualitative assessment of the quality of a measurement may add useful information for readers and for databases in guiding researchers to use the measurement appropriately in their analyses. Examples include indications of poor seeing or blended objects, uncertain deblending of spectral lines, redshifts based on a single spectral line assuming identification of the proper feature, etc.

### 2.6. *Classifications*

There are many spectral and morphological/phenomenological classifications for astronomical objects, and these classifications are constantly evolving. We suggest:

- **Utilize established classifications as available.** For basic morphological types, use the well-established schemes (e.g., Sandage 2005, and references therein). Authors are encouraged to refer to NED's extensive suite of searchable galaxy classifications and attributes (NED Team 2022)[17] or SIMBAD's Object Classification (SIMBAD Team 2021)[18], which have been standardized to enable unified queries across journal articles and catalogs.

- **Define new classifications clearly.** If new classifications must be created, define them clearly so they can be easily adopted by other researchers, and quickly integrated into databases.

### 2.7. *Orbital parameters*

Reports of measured orbital parameters often suffer from ambiguity in terminology used because the conventions used are not clear. It is advised to avoid ambiguous terms, and to explicitly define all terms and symbols wherever there is any possibility of confusion. For instance:

- **Avoid using the term "longitude of periapsis (or periastron)" when "argument of periapsis (or periastron)" is really the term intended.** Only use "longitude of periapsis" when referring to the sum of the argument of periapsis and the longitude of the ascending node.

- **Be explicit about which body's orbit a measured argument (or longitude) of periapsis refers to.** The argument of periapsis for a planet or a secondary star's orbit differs from that of the host or primary star's reflex motion by 180 degrees. Note also that the quantity is affected by the sign convention adopted in radial velocity analyses: as a general rule, positive radial velocity should indicate redshift, and negative should indicate blueshift.

- **Include time of periapsis as appropriate when reporting orbital elements.** For example, when reporting timing for a non-transiting eccentric orbit for which argument of periapsis is measured, report time of periapsis in preference to (or in addition to) time of inferior conjunction. Both are preferred if possible.

## 3. DATA PRESENTATION

### 3.1. *Tables*

Observational data are often published in the form of tables in an article. For tabular data presentation:

- **Provide a clear title and unambiguous labels of columns.** Indicate the units for each column when applicable. This is especially important when data are to be compared, or used outside a specialized field.

- **Explain the content of each column.** Clarify all special symbols or flags in a column (e.g., make a clear distinction between z the redshift and z the filter), and give references to cited values.

- **Keep each column homogeneous.** A single column should not present measurements with different units, mix errors with limits or comments, or append flags to values (see, e.g., the left hand side of Table 2).

- **Use clearly defined, non-numeric representations for missing (null) values.** It is recommended to use null values that are supported and documented by widely-used toolkits, e.g., "NaN" (Not a Number) for floating-point data in Astropy.[19] Avoid using numerical values, as they could be misinterpreted as actual measurements by people or software. For example, a computer will not hesitate to calculate the average of a list of redshift values that include "-999" or "0.0", deriving a misleading result, but it will properly ignore or issue a warning when "NaN" is encountered.[20] Use the same representation for missing data and have a separate field that explains the reasons for a missing value. Do not use different representations to indicate the different reasons, e.g., blank for "not observed", and "NaN" for "no detection".

- **Accompany a machine-readable table (MRT) with a ReadMe file.** Authors should include a human-readable description of the data, with at least the column descriptions, units, and references (on the origin of the measurements or instruments for observations when relevant) in a ReadMe file. A comprehensive example of the contents and structure of a ReadMe file can be found in the CDS document "Astronomical Catalogues and Tables Adopted Standards" (Ochsenbein 2000).[21] Authors are encouraged to utilize existing Python packages (e.g., cdspyreadme)[22] to generate ReadMe or MRTs from a wide list of standard table formats.

For tables containing astronomical objects, it is preferred that authors give the complete names of the objects (§2.1) in each table, and keep the same names in all the tables and text throughout the article when possible. This will not only enable the readers to quickly access related data about the objects, but also greatly expedite the linking and ingestion of data into data archives. The coordinates of the objects (§2.2) should be given in the table where the objects are first presented.

We recommend not to use LaTeX for large tables as the markups degrade the reusability of that data. Authors can improve the transfer of results by avoiding elaborate LaTeX tricks and treating inline tables as regular data arrays. For instance, method or other qualifying note marks could be given numerical values and displayed in an additional column instead of using table note marks which are not easily parsed by a machine (see Table 2). Additional details of the table should be included in an accompanying metadata file (see §4). Authors are also advised to report important numerical results for more than a couple of data points in tables, not just in text. When a number is reported in both text and a table, make sure it is consistent in both places.

### 3.2. *Figures*

Figures are commonly used for data visualization in an astronomical article, especially for images, spectra, SED plots, etc. For figures in a publication, we recommend:

---

[19] https://www.astropy.org/

[20] Sophisticated techniques for handling missing values for different data types in Python are described in many books and websites (e.g., McKinney 2017).

[21] https://vizier.unistra.fr/vizier/doc/catstd.htx

[22] https://github.com/cds-astro/cds.pyreadme; Landais (2016)

**Table 2**: Avoid LaTeX markups in tables to enhance reusability

| Object | Redshift |
| --- | --- |
| Source 1 | 0.1* |
| Source 2 | 0.2† |
| Source 3 | 0.3* |
| Source 4 | 0.4* |
| Source 5 | 0.5† |

| Object | Redshift | Quality |
| --- | --- | --- |
| Source 1 | 0.1 | 1 |
| Source 2 | 0.2 | 2 |
| Source 3 | 0.3 | 1 |
| Source 4 | 0.4 | 1 |
| Source 5 | 0.5 | 2 |

(a) Redshift quality flag: * = secure, † = uncertain.  (b) Redshift quality flag: 1 = secure, 2 = uncertain.

Note: The quality of the measurement is indicated using LaTeX markups (asterisks and daggers) on the left, and using an extra column with numerical values on the right. The table on the right is as human readable as the table on the left, but is more machine readable therefore making the data more reusable.

- **Provide clear caption, legend and axis labels for each figure.** Describe in detail what is presented in the figure, what different colors, symbols, and lines represent. Units of the axis labels should be included when applicable. In practice, figures should be able to stand alone without requiring much reading of the main text.

- **Create the graphics with accessibility in mind.** Accessibility should be taken into consideration when designing the graphics as it will help a broader audience to properly understand the information conveyed in these plots. For example, color-blind users would benefit from symbols that vary in shape in addition to colors. See the AAS[23] journals' graphics guide[24] for more advice on this.

- **Include information for "data behind the plots".** Make the original data files used to generate the figures publicly available, as this will greatly enhance the ability to reproduce, validate, or build upon published results.

We include an example of a clearly labeled, accessible plot in Figure 1. This was revised from a previously published article (Cook et al. 2019) with the first author's permission. The published version of this figure, although very accessible with information conveyed using variations in both colors and symbols/line styles, is lacking the units for the axis labels, and missing the explanation for some of the lines in the plot. The revision has addressed these issues.

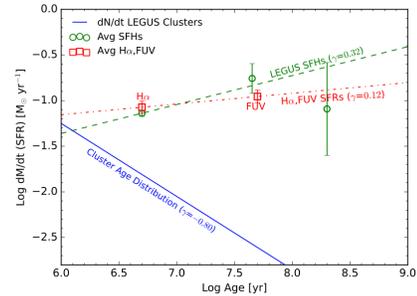

**The total SFR versus age of the LEGUS dwarf galaxies using two independent SFR measurements.**
The red squares represent the summed Hα and FUV SFRs corrected for dust extinction from Lee et al. (2009), we have updated the SFR conversion using the prescription of Murphy et al. (2011) with a Kroupa IMF. The green circles represent the summed SFRs derived from the resolved-star SFHs of Cignoni et al. (2018). The SFRs are preliminary and are in the process of being updated. The dashed green and dash-dot red lines represent linear fits to the Hα/FUV and SFH data points, respectively. The blue solid line represents the cluster age distribution in the LEGUS dwarf sample that has been scaled to fit on this graph for purposes of comparing the slopes. The total SFRs from both methods show a constant star formation in the composite dwarf sample indicating that the decline in the cluster age distribution is dominated by cluster disruption.

**Figure 1**: Example of a clearly labeled, accessible plot. We revised Figure 28 from Cook et al. (2019) with the help of the original lead author, adding the previously missed units for the axis labels and explanation of what each line is representing in the caption.

## 4. DATA ARCHIVING AND ACCESS

To ensure long term preservation, access, and reuse of data published in the literature, data products should be made available to readers and for data integration services. This applies to both data from original observations and enhanced products derived from previously published and/or archival data. We ask authors to:

- **Append small data sets as part of the publication.** Preserve data as supplementary materials with your final journal article, or post the data files with your arXiv preprint.

- **Deposit large or complex data at well-established data centers.** If the data are either too large or too complex to be hosted by the journal, authors are encouraged to place their data in a trusted repository that issues Digital Object Identifiers (DOIs, more on this in §5.4). Adhere to the specific format requirements from the archives when contributing data.

- **Provide a complete list of metadata.** Metadata provide valuable information on various aspects of the data such as the creation, standards and quality of the data, and are often necessary to fully comprehend the data. For example, to visualize the position and orientation of the apertures on imagery, key metadata including aperture dimensions, center coordinates, and position angle are required. For reference, MAST[25] has compiled a comprehensive list of the required, recommended and suggested metadata common to different types of astronomical science data.[26]

---

[23] The American Astronomical Society; https://aas.org/

[24] https://journals.aas.org/graphics-guide/#preparing_files

- **Include a "Data Availability Statement" if required by the journal.** The statement provides a standardized format for the transparent reporting and description of the data underlying the article. For example, the Monthly Notices of the Royal Astronomical Society (MNRAS) requires all authors to include a "Data Availability Statement" in their articles. The requirement and many sample statements can be found in its "Instructions to Authors".[27]

- **Do not publish data sets at URLs lacking long-term support.** We strongly discourage the publication of URLs to personal web servers hosting data sets for which the author or institution has no means to maintain for many years after the publication of the associated journal article. This will greatly hinder future effort by researchers to retrieve and reuse the data.

When depositing file-based data, it is recommended to use unique and informative names for the files instead of duplicating file names and using location in a directory structure as file metadata necessary to uniquely identify a file. For example, if photometry data are available at different bands (e.g., V and R) for the same object (e.g., NGC 1275), use names such as NGC1275_V.dat and NGC1275_R.dat to identify the files. Do not set up separate directories for V and R band, and give the same file name NGC1275.dat under both directories.

In Appendix B, we provide a partial list of well-reputed data repositories that support long-term storage and access to scientific catalogs, images, spectra, light curves, data cubes, radio visibilities, probabilistic density functions and other large data files.

## 5. CITATIONS AND CREDITS

### 5.1. *Literature citations*

The AAS president in May/June 2009, Dr. John Huchra, wrote: "Authors have a strong tendency to under-attribute, that is not to properly cite both previous ideas and basic data. The phrases 'data are available' or 'it is known that' should never be used; someone had to work fairly hard to get that data or develop that idea and they deserve credit, even if they are your competitors. Also complicated is the use of master compilations, easy, but again not giving credit where credit is really due."[28]

To properly cite data in an article, one should always:

- **Cite the original references.** For example, if the redshift presented is obtained from NED, use phrases such as "We adopted a heliocentric redshift of 1.234 (Smith et al. 2012) via NED", where "Smith et al. 2012" is listed correctly in your bibliography.

- **Use preferred citations by the authors.** Follow the instructions provided by the authors of the original papers, software etc. and use what the authors preferred for citations. For example, if 2MASS data are used in your analysis, the 2MASS web page[29] requests that you cite the canonical paper by Skrutskie et al. (2006), instead of the Explanatory Supplement.[30] A separate standard acknowledgment is also listed there to be included in the acknowledgments.

- **Provide full provenance of the data.** Ensure full reproducibility of the analysis performed in the paper by citing all data/source/software used: explicitly list the data identifiers provided by the data center in a table or in the text. For example, if 2MASS data is used via TOPCAT accessing the VizieR table, then in addition to any preferred citations of the data itself, cite the software and compilation used, e.g.,"2MASS (Skrutskie et al. 2006) as downloaded with TOPCAT (Version 4.8-3, Taylor 2005) via VizieR (II/246, Cutri et al. 2003)". It is also recommended to include the names of principal investigators who acquired the original data sets.

- **Include all references in the bibliography.** Make sure all appropriate references to papers, software and data products are included in a paper's bibliography section, not just in footnotes.

---

This will ensure proper attribution of citations to them. More details and examples on this can be found in the AAS Journal Reference Instructions.[31]

- **Distinguish original data in your article and data taken from other work.** Use phrases such as "This work" to clearly identify original data in your article. It is important to archives, data services, and other authors that papers distinguish new information and measurements from those made by previous work.

### 5.2. *Facility credits*

Proper credit needs to be given to the facility or service used to obtain the data in an article. This may be hardware (telescopes and instruments) or on-line services (databases). Not only are the specifics essential to properly understand the data, the acknowledgement metric is also used as a basis for productivity by organizations maintaining telescopes, archives serving the data, and funding agencies for those facilities. We therefore recommend:

- **Explicitly indicate the facility involved.** Always describe the facilities or services used, and make sure the name is unique. Examples of some of the ambiguous names seen in the literature are given in Table 3.

- **Use standardized keywords when possible.** The AAS has created keyword tags[32] to be used with AASTeX \facility and \facilities. Many major astronomical archives, databases, and computational centers have been recently added to this list (e.g., CDS, Exoplanet Archive, IRSA[33], NED).

- **Include facility's own statement if available.** Refer to the phrase that was in place at the time the facility was used. For example, the NASA Exoplanet Archive asks authors to include the following standard acknowledgment in any published material that makes use of its services: "This research has made use of the NASA Exoplanet Archive, which is operated by the California Institute of Technology, under contract with the Na-

tional Aeronautics and Space Administration under the Exoplanet Exploration Program." [34]

### 5.3. *Software credits*

The increasing volume and complexity of the astronomical data require a lot of software processing in obtaining, calibrating, and analyzing the data and building the final product. Similar to acknowledging the facilities and services, we recommend:

- **List the software and version used in the production of the article.** Use the preferred citation if available, e.g., the paper describing the software. If not, include the name of the author(s), title of the program/source code, the code version and a URL link to the code publisher. For example, "Stanford Classifier v3.9.2, The Stanford Natural Language Processing Group, https://nlp.stanford.edu/software/classifier.shtml" is in the Acknowledgment section in Chen et al. (2022). This will ensure proper credit is given, and at the same time help with reproducibility.

The Astrophysics Source Code Library (ASCL)[35] is a free registry for authors to submit the software used in their research that has appeared in or been submitted to peer-reviewed journals. A unique ASCL ID is assigned to each code which enables accurate citation.

### 5.4. *Digital object identifiers*

A digital object identifier (DOI)[36] provides a persistent and unique identification of online resources, and is widely used to identify content related to published articles. Authors are encouraged to:

- **Use DOIs to cite related content if available.** This includes specific data sets, software, and services used to produce results in the published articles. Archive these in persistent repositories and link them to the article through DOIs minted by the repositories. The DOI links should be included in the bibliography to ensure proper citation (see the bibliography section of this article for examples), and also be put where the data are discussed to make it easier for readers to locate and access the data.

Some resources for obtaining DOIs for specific data sets at the archives, as well as guidelines for using them

---

**Table 3**: Examples of ambiguous facility/telescope/instrument names in literature

| Name as published | Possible interpretation |
| --- | --- |
| ARO | Astronomical Research Observatory |
| | Arizona Radio Observatory |
| | Abbey Ridge Observatory |
| | Algonquin Radio Observatory |
| DDO | David Dunlap Observatory:0.15m |
| | David Dunlap Observatory:0.5m |
| | David Dunlap Observatory:0.6m |
| | David Dunlap Observatory:1.88m |
| EMIR | Eight MIxer Receiver (on the IRAM 30m radio telescope) |
| | Espectrógrafo Multiobjeto Infra-Rojo (on the Gran Telescopio Canarias) |
| OSIRIS | OH-Suppressing Infra-Red Imaging Spectrograph (on the Keck I telescope) |
| | Ohio State Infrared Imager/Spectrograph (on the SOAR telescope) |
| | Optical System for Imaging and low-Intermediate-Resolution Integrated Spectroscopy (on the Gran Telescopio Canarias) |

in the journal articles, are provided in Appendix C. When DOIs are not available, authors can use the exact data set identifiers from the archives where the data were obtained. For example, the Spitzer Survey of Stellar Structure in Galaxies (S4G) dataset in IRSA was cited (by footnote) as http://irsa.ipac.caltech.edu/data/SPITZER/S4G/ in Ciambur (2015) before the DOI for this dataset was generated. Better reference methods currently include citing this dataset URL in the bibliography, e.g., S4G Team (2022), or citing the specific IRSA S4G DOI, S4G Team (2020).

## 6. DATA CONTENT KEYWORDS

Traditional keywords used in journal articles do not capture information about the types of data presented in the article, for example, whether the article presents new position measurements (astrometry), photometric data, spectroscopic data, etc. This greatly limits the possibility of filtering literature searches based on the availability of these specific types of data. We suggest that authors:

- **Tag articles with relevant data content keywords from the Unified Astronomy Thesaurus (UAT).**[37] The UAT is an open and community-supported project that formalizes astronomical concepts (Frey & Accomazzi 2018). It is adopted as a standard by the AAS journals[38]

---

[37] https://astrothesaurus.org/

[38] https://journals.aas.org/aas-journals-uat/

---

and the broader astronomical community, including ADS, to tag astronomical work with accurate, broadly adopted concepts. Hence we recommend that authors tag their articles with UAT keywords that best describe the types of data contained in the article. The keywords for this article are defined using the UAT.

## 7. SUMMARY

We hope the guidelines provided here will assist authors in preparing and publishing their data in a way that allows science claims to be clearly communicated, and also readily understood and validated by readers. These best practices are intended to be used not only by authors during the preparation and submission stages of a publication, but also by referees, editors and publishers during the refereeing and editing stages before final publication. This will improve the quality of the published research record, expedite the integration of data into the databases with more efficiency and accuracy, and result in long-term preservation and reuse of valuable data. This will in turn enable more scientific discoveries that would otherwise not be possible or practical, and increase citations for authors. A copy of this document and the checklist can be accessed at https://ned.ipac.caltech.edu/uri/Docs::BPDP. We expect that these best practices will evolve as new forms of data are published in the journals, and we welcome feedback and input from authors, publishers, archive users, and all interested parties.

ACKNOWLEDGMENTS

We thank Frank Timmes from Arizona State University for his helpful comments to improve this work. We appreciate the endorsement from the following colleagues:

1. Kim Clube, Royal Astronomical Society, UK

2. Uta Grothkopf, European Southern Observatory, Germany

3. Mark Lacy, National Radio Astronomy Observatory, USA

4. Adam Leary, Oxford University Press, UK

5. Greg J. Schwarz, American Astronomical Society, USA

# APPENDIX

## A. CHECKLIST OF RECOMMENDATIONS FOR PUBLISHING DATA IN THE LITERATURE

This checklist is a digest version of detailed information given in this article. It is intended to be a short reference for authors, referees, and science editors to consult in order to avoid various pit-falls that often impede the interpretation of data and metadata by readers, and parsing by software, and therefore also complicate and delay integration of the data into astronomical databases.

1. General rules (§2)

   (a) Define all symbols, acronyms, and abbreviations at first use.

   (b) Provide uncertainty and confidence level when reporting a new measurement.

   (c) Present the appropriate number of significant figures for numerical measurements and uncertainties that match the precision of the measurements.

   (d) Report the units for measurements if present, and adopt commonly-used ones.

   (e) Indicate preferred values if applicable.

2. Nomenclature (§2.1)

   (a) Provide the complete name for each object. (§2.1.1)

   (b) Include the "J" in names based on J2000 coordinates. (§2.1.1)

   (c) Insert spacers between a catalog name and the identifiers within the catalog. (§2.1.1)

   (d) Distinguish between part of an object and the object itself. (§2.1.1)

   (e) Do not use the same name for different objects. (§2.1.1)

   (f) Always assign a name and verify the name is unique. (§2.1.2)

   (g) Keep the appropriate number of significant figures in coordinate-based names. (§2.1.2)

   (h) Use established names for known objects and check for the correct formatting. (§2.1.3)

   (i) Confirm the names and positions for cross-identifications. (§2.1.4)

   (j) Cross-match the same objects in different tables within the same article. (§2.1.4)

3. Astrometry (§2.2)

    (a) Provide the best available coordinates.

    (b) Specify the celestial reference system and/or frame.

    (c) Indicate the equinox and epoch of observation when necessary.

    (d) State the wavelength range from which astrometry is obtained.

4. Photometry (§2.3)

    (a) State the facility, telescope and instrument used.

    (b) Describe the method used to estimate photometry.

    (c) Use standard passband/filter identifiers.

    (d) Clarify the magnitude system.

    (e) Specify spectral transitions completely.

5. Time (§2.4)

    (a) Provide the time of observation and exposure time.

    (b) Favor full Julian Dates over abbreviated or offset Julian Dates.

    (c) Include phase timing measures along with reported periods when relevant.

    (d) State when observations from multiple missions are executed simultaneously.

6. Redshift/velocity (§2.5)

    (a) Describe the method of redshift measurements (spectroscopic, photometric, etc.) and give references to the model/method.

    (b) Specify the reference frame of the redshift measurements (barycentric, heliocentric, galactocentric, etc.).

    (c) Provide the frequency/wavelength from which the measurement is obtained.

    (d) State whether a published recessional velocity is based on observed frequency or wavelength shifts (i.e., radio or optical convention).

    (e) Indicate the quality of the measurement when possible.

7. Classifications (§2.6)

    (a) Utilize established classifications as available.

    (b) Define new classifications clearly.

8. Orbital parameters (§2.7)

    (a) Avoid using "longitude of periapsis" in place of "argument of periapsis".

    (b) Be explicit about which body's orbit a longitude or argument of periapsis refers to (e.g., planet or host star).

    (c) Include time of periapsis as appropriate.

9. Tables (§3.1)

    (a) Provide a clear title and unambiguous labels for columns.

    (b) Explain the content of each column, including symbols and flags.

    (c) Keep each column homogeneous.

    (d) Use the same explicitly defined non-numeric representations for missing (null) values throughout.

    (e) Prepare ReadMe files for machine-readable tables.

10. Figures (§3.2)

    (a) Provide clear caption, legend and axis labels for each figure.

    (b) Design the graphics to be accessible.

    (c) Make "data behind the plots" publicly available.

11. Data archiving and access (§4)

    (a) Append small data sets as part of the publication.

    (b) Deposit large or complex data at a long-term archive most appropriate for your data. Adhere to the specific format requirements from the archives.

    (c) Provide a complete list of metadata.

    (d) Include a Data Availability Statement if required by the journal.

    (e) Do not publish data sets at URLs lacking long-term support.

12. Literature citations (§5.1)

    (a) Cite the original references.

    (b) Use preferred citations by the authors.

    (c) Provide full provenance of the data. Credit the originator of archival data, including the Principal Investigator.

    (d) Include all references in the bibliography section.

    (e) Distinguish original data in your article and data taken from other work.

13. Facility credits (§5.2)

    (a) Indicate the facilities involved, such as telescopes, instruments, and databases.

    (b) Use standard keywords when possible.

    (c) Include facility's own statement if available.

14. Software credits (§5.3)

    (a) List the software and version used in the production of the article.

15. Digital object identifiers (§5.4)

    (a) Use DOIs to cite data sets, software and services if available.

16. Data content keywords (§6)

    (a) Tag articles with relevant data content keywords from the UAT.

## B. DATA REPOSITORIES

We list here some data repositories (both special-purpose and general-purpose ones) that are available to support long-term storage and access to scientific catalogs, images, spectra, light curves, code, and other large data files used to produce the figures and results that appear in journal articles. This is not a complete list. Authors are encouraged to explore and choose a well-established repository that is most appropriate for your data. At the time your data are contributed, please acquire a DOI or other persistent URL suitable for publication in the journal article.

**Astronomy data repositories:**

1. CANFAR Data Publication Service; https://www.canfar.net/en/docs/digital_object_identifiers/; data, figures, software, or any other material that is important to an astrophysics publication.

2. ESO Science Archive; https://archive.eso.org/cms.html; astrophysics data and metadata.

3. ExoFOP; https://exofop.ipac.caltech.edu/; online environment for the community to upload and share observations, data, files, and notes on exoplanet candidates.

4. HEASARC; https://heasarc.gsfc.nasa.gov/; astrophysics catalogs, images, spectra, time series.

5. IRSA; https://irsa.ipac.caltech.edu/frontpage/; astrophysics catalogs, images, spectra, time series.

6. KOA; https://koa.ipac.caltech.edu; astrophysics catalogs, images, spectra, time series; raw and contributed science products.

7. MAST; https://archive.stsci.edu/; astrophysics catalogs, images, spectra, time series, publication records, etc.

8. NASA Exoplanet Archive; https://exoplanetarchive.ipac.caltech.edu/; data and derived astrophysical parameters for exoplanets and exoplanet candidates.

9. PaperData; https://paperdata.china-vo.org/; astrophysics data, software, code, and metadata.

10. VizieR; https://vizier.unistra.fr/; astrophysics catalogs, images, spectra.

**General-purpose data repositories:**

1. Code Ocean; https://codeocean.com/; all research data and code.

2. DataHub; https://datahub.io/; all research data and metadata.

3. Figshare; https://figshare.com/; all research data and metadata.

4. Harvard Dataverse; https://dataverse.harvard.edu/; all research data and metadata.

5. Mendeley Data; https://data.mendeley.com/; all research data and metadata.

6. Open Science Framework; https://osf.io/; all research data and metadata.

7. ScienceDB; https://www.scidb.cn/en; all research data and metadata.

8. Zenodo; https://zenodo.org/; all research data and metadata.

## C. LIST OF DOIS

**Resources for publishing DOIs in astrophysics journals:**

1. AAS Journals guidelines for citing 3rd party data repositories and software; https://journals.aas.org/aastexguide/#softwareandthirdparty.

2. A&A data policy on publishing and archiving the data, https://www.aanda.org/for-authors/author-information/paper-organization#Data%20policy

3. MNRAS data policy on availability and citation of data; https://academic.oup.com/mnras/pages/General_Instructions#2.9%20Data%20Policy.

4. Oxford University Press guidelines for selecting a repository that will issue a DOI; https://academic.oup.com/journals/pages/authors/preparing_your_manuscript/research-data-policy#choosing.

**Resources for obtaining DOIs from astrophysics archives:**

1. CDS; CDS creates DOIs for catalogs when they are ingested in VizieR. This DOI is for the digital object of the catalog in VizieR and is complementary to the DOI of the publication.

2. Chandra Data Archive; https://cxc.cfa.harvard.edu/cda/cda_doi.html.

3. Exoplanet Follow-up Observing Program (ExoFOP);https://www.ipac.caltech.edu/dois/exofop.

4. IPAC Archives; https://www.ipac.caltech.edu/dois/.

5. IRSA; https://www.ipac.caltech.edu/dois/irsa.

6. KOA; https://www.ipac.caltech.edu/dois/koa.

7. MAST; https://archive.stsci.edu/publishing/doi.

8. NASA Exoplanet Archive; https://www.ipac.caltech.edu/dois/exoplanet-archive.

9. NED; https://www.ipac.caltech.edu/dois/ned.